\documentclass[osajnl2,superscriptaddress,twocolumn]{revtex4}
\usepackage{mathptmx}
\usepackage[T1]{fontenc}
\usepackage[latin9]{inputenc}
\usepackage{verbatim}
\usepackage{booktabs}
\usepackage{amsmath}
\usepackage{graphicx}
\usepackage{amssymb}
\usepackage{esint}

\makeatletter

\providecommand{\tabularnewline}{\\}

\@ifundefined{textcolor}{}
{%
 \definecolor{BLACK}{gray}{0}
 \definecolor{WHITE}{gray}{1}
 \definecolor{RED}{rgb}{1,0,0}
 \definecolor{GREEN}{rgb}{0,1,0}
 \definecolor{BLUE}{rgb}{0,0,1}
 \definecolor{CYAN}{cmyk}{1,0,0,0}
 \definecolor{MAGENTA}{cmyk}{0,1,0,0}
 \definecolor{YELLOW}{cmyk}{0,0,1,0}
 }

\usepackage[titles]{tocloft}
\renewcommand\cftdotsep\cftnodots
\cftsetindents{figure}{0em}{1.5em}
\cftpagenumbersoff{figure}

\makeatother

\begin{document}

\title{Optical memory based on ultrafast wavelength switching in a bistable
microlaser}

\author{Sergei V. Zhukovsky}

\affiliation{Theoretical Nano-Photonics Group, Institute of High-Frequency and
Communication Technology, Faculty of Electrical, Information and Media
Engineering, University of Wuppertal, \\
Rainer-Gruenter-Str.~21, D-42119 Wuppertal, Germany.}

\altaffiliation{sergei@uni-wuppertal.de, chigrin@uni-wuppertal.de}

\author{Dmitry N. Chigrin}

\affiliation{Theoretical Nano-Photonics Group, Institute of High-Frequency and
Communication Technology, Faculty of Electrical, Information and Media
Engineering, University of Wuppertal, \\
Rainer-Gruenter-Str.~21, D-42119 Wuppertal, Germany.}
\begin{abstract}
We propose an optical memory cell based on ultrafast wavelength switching
in coupled-cavity microlasers, featuring bistability between modes
separated by several nanometers. A numerical implementation is demonstrated
by simulating a two-dimensional photonic crystal microlaser. Switching
times of less than 10~ps, switching energy around 15--30~fJ and
on-off contrast of more than 40~dB are achieved. Theoretical guidelines
for optimizing the performance of the memory cell in terms of switching
time and energy are drawn.
\end{abstract}

\pacs{210.4680, 130.4815, 190.1450, 140.3948, 230.5298}

\maketitle
In recent years, several designs of an optical memory (flip-flop)
cell were proposed. Elements based on bistability in active laser-based
systems \cite{KawaguchiReview,OksanenAPL,LiuIEEE,OsborneOE} %
\begin{comment}
and passive nonlinear \cite{NLAndabibJOSA,NLTanabeOL} 
\end{comment}
{}were shown to be promising in terms of switching performance. %
\begin{comment}
Laser-based systems can be superior to the passive nonlinear ones
due to the absence of a high-power holding optical beam in the cell
design \cite{KawaguchiReview}. 
\end{comment}
{}All-optical switching between the two memory states at the sub-nanosecond
time-scale was reported for directionally bistable injection-locked
microring lasers \cite{HillNature,LiuIEEE} and for bistable gain-quenching
semiconductor optical amplifiers \cite{OksanenAPL}. Microlasers offer
an advantage in the switching energy due to a small cavity size and
high Q-factors \cite{ourPRA09}. Recently, it was reported \cite{ourPSS07,ourPRL07,ourPRA09}
that a coupled-cavity microlaser can lase into either of its supermodes
in a bistable manner with a possibility of ultrafast picosecond-scale
switching between these modes \cite{ourPRL07}. %
\begin{comment}
The laser remains constantly above threshold wintout any pump modulation,
which is favorable for fast switching \cite{OksanenAPL}.
\end{comment}
{}

\begin{figure}[b]
\includegraphics[width=0.9\columnwidth]{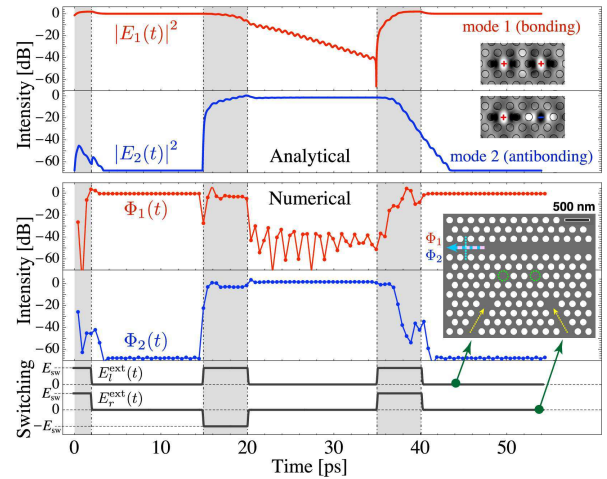}

\caption{(Color online) Analytical and numerical demonstration of a $1\to2\to1$
mode switching in a 2D PhC coupled-defect microlaser. $E_{_{l,r}}^{\text{ext}}(t)$
are switching signals in the defects; $\Phi_{1,2}(t)$ are FDTD output
for the two modes; $E_{j}(t)$ are obtained from Eqs.~\eqref{eq:Eb}--\eqref{eq:Wb}.
\label{fig:numerical}}

\end{figure}

In this Letter, we build up on our recent findings on bistability
in microlasers \cite{ourPSS07,ourPRL07,ourPRA09} to propose a design
of an ultrafast optical memory cell. We consider two identical coupled
defects in a two-dimensional (2D) triangular protonic crystal (PhC)
lattice (see Fig.~\ref{fig:numerical}) with active centers (e.g.,
quantum dots or quantum wires) embedded into the defects \cite{AtlasovAPL07,AtlasovAPL08}.
Such structures are within the state of the art of current fabrication
possibilities \cite{AtlasovOE}. Direct SNOM observations confirm
the formation of coupled-cavity supermodes in such a system \cite{IntontiSNOM}.
Unlike in polarization bistability, directional bistability in bidirectional
ring lasers, or bistability among longitudinal modes in unidirectional
ring lasers \cite{NarducciPRA}, bistable modes in coupled-cavity
microlasers can be up to tens of nanometers apart from each other
\cite{ourPRL07}, which makes the output signal subject to an easy
spectral separation compatible with a WDM arrangement. Our numerical
simulations show that for a laser operating 100--300 times above threshold
(which is still practically realizable), switching between modes 3~nm
apart can occur at less than 10~ps (see Fig.~\ref{fig:numerical}),
with switching energy estimated to be around 15--30~fJ.

\begin{figure}[b]
\includegraphics[width=1\columnwidth]{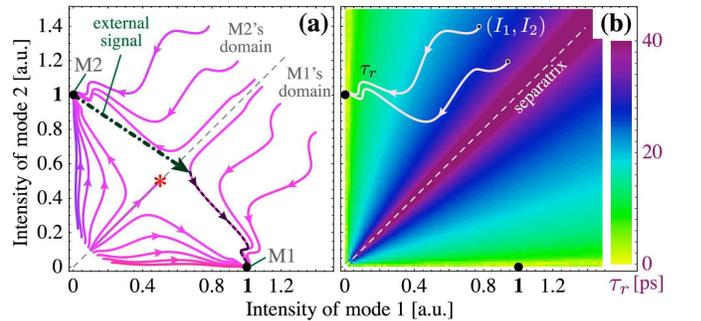}\caption{(Color online) (a) An above-threshold flow diagram for class-B two-mode
microlaser described by Eqs.~\eqref{eq:Eb}--\eqref{eq:Wb} showing
bistable lasing regime and the schematics of the mode switching; (b)
density plot of time~$\tau_{r}(I_{1},I_{2})$ it takes the system
to relax from the point of origin $(I_{1},I_{2})$ to the 20~dB-neighborhood
of the corresponding stable point. \label{fig:returnmap}}

\end{figure}

To describe the laser dynamics, we use the coupled mode theory \cite{ourPRA09},
which is a generalization of existing models (see, e.g., \cite{NarducciPRA})
for the case of arbitrary cavity shape. A resonator comprising two
identical coupled cavities supports two modes (the bonding and the
antibonding supermodes) characterized by symmetric (in-phase) vs.~antisymmetric
(out-of-phase) cavity field distribution, respectively (see inset
in Fig.~\ref{fig:numerical}) We assume the supermodes to have frequencies~$\omega_{1,2}=\omega_{0}\mp\Delta_{\omega}$
and decay rates~$\kappa_{1,2}$. The total cavity field $E(\mathbf{r},t)$
can be written in terms of spatial mode distributions~$u_{j}(\mathbf{r})$
and time-dependent mode amplitudes~$E_{j}(t)$ as $E(\mathbf{r},t)=u_{1}(\mathbf{r})E_{1}(t)\mathrm{e}^{-i\omega_{1}t}+u_{2}(\mathbf{r})E_{2}(t)\mathrm{e}^{-i\omega_{2}t}$.
In the limit of class-B laser, the mode amplitudes obey the equations
\cite{ourPRA09}\begin{equation}
\dot{E}_{j}=-(\kappa_{j}/2)\left(E_{j}+E_{j}^{\text{ext}}\right)+(\zeta/\beta_{j})\sum_{k}E_{k}N_{jk}\mathrm{e}^{i(\omega_{j}-\omega_{k})t},\label{eq:Eb}\end{equation}
where $\beta_{1,2}$=$\Delta_{a}\pm i\Delta_{\omega}$ and $\zeta\approx\mu^{2}\omega_{0}/2\epsilon_{0}\epsilon\hbar$.
The terms~$E_{j}^{\text{ext}}(t)$ account for the $j$-th mode projection
of the externally injected signal \cite{ourPSSMethods07}. $N_{jk}(t)=\epsilon\int\mathrm{d}^{3}\mathbf{r}\, u_{j}^{*}(\mathbf{r})N(\mathbf{r},t)u_{k}(\mathbf{r})$
are projections of population inversion~$N(\mathbf{r},t)$ governed
by the equations \begin{equation}
\dot{N}_{jk}=\gamma\left(R_{jk}-N_{jk}\right)-\xi\sum_{mnpq}\mathbb{M}_{jk}^{mn,pq}N_{pq}E_{m}^{*}E_{n}\mathrm{e}^{i(\omega_{m}-\omega_{n})t}.\label{eq:Wb}\end{equation}
Here $\xi=\mu^{2}/4\hbar^{2}$ and~$R_{jk}$ are projections of the
external pumping rate~$R(\mathbf{r},t)=R$. Other parameters are
listed in Table~\ref{tab:parameters}%
\begin{comment}
(see also Ref.~\cite{ourPRA09} for further details)
\end{comment}
{}. The elements of the matrix~$\mathbb{M}$ are composed of $\beta_{j}$
and $\alpha_{jk}^{mn}=\epsilon\int\mathrm{d}^{3}\mathbf{r}\, u_{j}^{*}(\mathbf{r})u_{k}(\mathbf{r})u_{m}^{*}(\mathbf{r})u_{n}(\mathbf{r})$.
For symmetric pumping and near-orthogonal modes, the non-zero elements
are \begin{equation}
\begin{gathered}\mathbb{M}_{jj}^{mm,pp}=\alpha_{jj}^{mm}\left(1/\beta_{m}+1/\beta_{m}^{*}\right),\quad\mathbb{M}_{jk}^{mm,pq}=\alpha_{jk}^{qp}/\beta_{p},\\
\mathbb{M}_{jj}^{mn,mn}=\left(\alpha_{jj}^{mm}+\alpha_{jj}^{nn}\right)/\beta_{m},\quad\mathbb{M}_{jk}^{mn,pp}=\alpha_{jk}^{mn}/\beta_{n}.\end{gathered}
\label{eq:M}\end{equation}
It was shown that Eqs.~\eqref{eq:Eb}--\eqref{eq:Wb} support bistable
solutions provided $\alpha_{kk}^{jj}\simeq\alpha_{jj}^{jj}$ and the
pumping rate is high enough \cite{ourPRA09}. A corresponding flow
diagram is shown in Fig.~\ref{fig:returnmap}a, clearly indicating
two stable states (M1 and M2) with two equally sized domains of attraction
divided by a separatrix.

\begin{table}
\caption{The parameter values used in the calculations.\label{tab:parameters}}

\begin{tabular}{cll}
\toprule 
Symbol & Value & Comment (see also \cite{ourPRA09} for details)\tabularnewline
\midrule
$\mu$ & $\sim10^{-28}\,\mathrm{A}\cdot\mathrm{s}$ & dipole moment of atomic laser transition\tabularnewline
$\epsilon$ & $12.3763$ & dielectric constant of the host (GaAs)\tabularnewline
$\omega_{0}$ & $2.13\times10^{15}\,\mathrm{s}^{-1}$ & operating frequency (around 885~nm)\tabularnewline
$\Delta_{a}$ & $1.88\times10^{13}\,\mathrm{s}^{-1}$ & polarization decay rate (laser linewidth)\tabularnewline
$\Delta_{\omega}$ & $0.1\Delta_{a}$ & intermode frequency separation\tabularnewline
$\kappa_{j}$ & $0.1\Delta_{a}$ & cavity mode decay rate\tabularnewline
$\gamma$ & $0.001\Delta_{a}$ & population inversion decay rate\tabularnewline
$\alpha_{jk}^{mn}$ & $3.51\times10^{11}$ & non-zero mode overlap integrals\tabularnewline
\bottomrule
\end{tabular}

\end{table}

The principle of optical memory operation proposed here is based on
the switching of the bistable microlaser between its two stable states
by injection locking. Either mode can be locked into by choosing the
right spatial profile of the injection signal~$E_{j}^{\text{ext}}$
while keeping the same spectral profile \cite{ourPRL07}. Suppose
the laser is at M2, i.e., generating into mode~2 so that $E_{2}(t)=E_{\text{mode}}$.
An injected pulse shifts the mode balance in favor of mode~1, i.e.,
moves the resonator away from M2 and over the separatrix. Due to bistability,
the system will relax to the other stable point M1, i.e., ends up
lasing into mode~1 (see Fig.~\ref{fig:returnmap}a). In order to
reach the separatrix the minimal energy delivered by the switching
pulse $W_{\text{sw}}$ should be at least $W_{\text{c}}/\sqrt{2}$
where $W_{\text{c}}$ is the energy stored in the cavity.

In response to the switching pulse, assumed to be a step function
with height~$E_{\text{sw}}$ and duration~$\sigma$, two processes
can be distinguished (Fig.~\ref{fig:returnmap}a): evolution of the
system away from M2 under the action of the pulse and relaxation to
M1 after the pulse has subsided. The former is time-limited by the
switching pulse duration~$\sigma$. The duration of the latter~$\tau_{r}$
depends only on the initial point $(I_{1},I_{2})$ on the flow diagram
that the switching pulse brings the system to. This duration is large
near the separatrix and rapidly decreases away from it (Fig.~\ref{fig:returnmap}b).
It is thus desired that the switching pulse brings the system close
to mode-1 axis or M1. The desired pulse energy is then $W_{\text{sw}}\simeq W_{c}\sqrt{2}$.
When~$W_{\text{sw}}$ varies between $W_{c}/\sqrt{2}$ and $W_{c}\sqrt{2}$,
the switching time is expected to drop significantly.

\begin{figure}
\includegraphics[width=0.9\columnwidth]{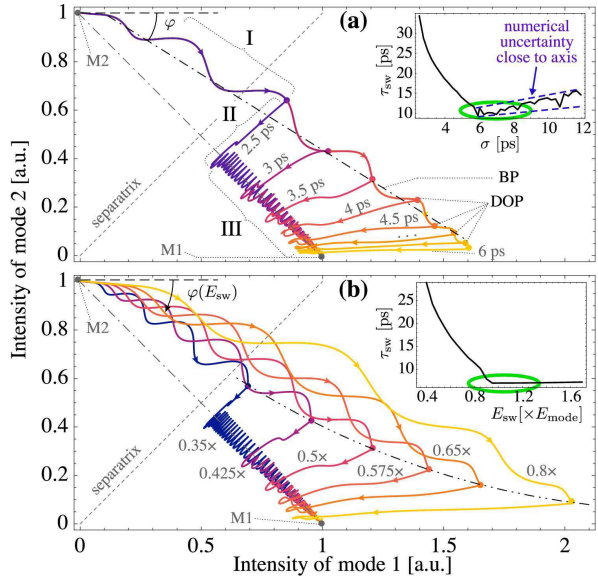}

\caption{(Color online) $(I_{1},I_{2})$-space diagrams of the $2\to1$ mode
switching by a stepwise pulse: (a) with fixed $E_{\text{sw}}=0.5E_{\text{mode}}$
and varying~$\sigma$; (b) with fixed $\sigma=3.5$~ps and varying
$E_{\text{sw}}$. The insets show the dependencies $\tau_{\text{sw}}(E_{\text{sw}})$
and $\tau_{\text{sw}}(\sigma)$. \label{fig:switchingpath}}

\end{figure}

This can be confirmed by plotting the switching diagrams in $(I_{1},I_{2})$-space
(Fig.~\ref{fig:switchingpath}). One can see that while the pulse
remains in effect, the system evolves along a straight line, \emph{build-up
path} (BP), accompanied by relaxation oscillations%
\begin{comment}
 caused by the injected pulse bringing the field--population inversion
($E$--$N$) coupling out of balance
\end{comment}
{}. The pulse duration~$\sigma$ determines how far the system progresses
along BP (I in Fig.~\ref{fig:switchingpath}a). As soon as the pulse
ends, the system leaves BP at a \emph{drop-off point} (DOP), quickly
relaxing onto the M2-M1 line (II) and then slowly continuing along
this line towards the other stable point (III). %
\begin{comment}
The latter process is again accompanied by relaxation oscillations.
\end{comment}
{}

The slope angle~$\varphi$ of the BP depends on the balance between~$E_{\text{sw}}$
and~$\kappa$. Higher $E_{\text{sw}}$ increases the injection rate
for the mode~1 while higher $Q$-factors (smaller~$\kappa$) make
the decay of the mode~2 slower. Both effects lead to a decrease of~$\varphi$
(Fig.~\ref{fig:switchingpath}b). Hence, setting the finesse of the
resonator too high makes switching excessively long while lowering
the $Q$-factors increases~$R_{\text{thr}}$ and, in turn, the operating
pumping rate~$R$. $E_{\text{sw}}$~and~$\sigma$ can be selected
so as to provide the fastest switching. Most of the BP should be covered
and the DOP should be as close to M1 as possible (Fig.~\ref{fig:switchingpath}).
There is a clear tradeoff between switching time and energy, as seen
in the plots $\tau_{\text{sw}}(E_{\text{sw}},\sigma)$ in Fig.~\ref{fig:switchingpath}.
A similar tradeoff was recently observed in low-energy all-optical
switching between transverse patterns in atomic vapor and semiconductor
microresonators \cite{Transverse08,Transverse09}. Further increase
of~$E_{\text{sw}}$ or~$\sigma$ moves the DOP away from M1, causing
a slight growth of~$\tau_{\text{sw}}$. Operating the laser beyond
the minimum in~$\tau_{\text{sw}}$ may be useful to improve the reliability
if system parameters vary due to fabrication disorder or environmental
factors.

These optimization guidelines were used in the design of a proof-of-the
principle memory cell for direct numerical simulations (Fig.~\ref{fig:numerical}).
The choice of geometry (coupled defects in a PhC lattice) was based
on state-of-the-art fabrication capabilities \cite{AtlasovOE}. The
active medium was modeled using 4-level laser rate equations coupled
to a 2D finite-difference time-domain (FDTD) solver \cite{ourPSSMethods07}.
Switching is effectuated by in-phase ($E_{l}^{\text{ext}}=E_{r}^{\text{ext}}$)
and out-of-phase ($E_{l}^{\text{ext}}=-E_{r}^{\text{ext}}$) quasi-monochromatic
rectangular pulses with frequency~$\omega_{0}$ delivered into the
defects through PhC waveguides. Due to symmetry matching, these patterns
couple to the bonding (M1) and the antibonding (M2) mode, respectively%
\begin{comment}
, without having to alter the temporal or spectral properties of the
switching pulses in any other way
\end{comment}
{}. The output radiation is extracted from another PhC waveguide (the
{}``bus''). The signal for the two frequencies is determined as
the across-the-bus flux of the Poynting vector $\Phi_{1,2}(t)\sim\int_{\text{bus}}\mathbf{E}(\mathbf{r},\omega_{1,2},t)\times\mathbf{B}^{*}(\mathbf{r},\omega_{1,2},t)\cdot\mathbf{n}\mathrm{d}^{2}\mathbf{r}$
obtained from coarse-graining Fourier transform of the fields at~$\omega_{1,2}$
as $\mathbf{E}(\mathbf{r},\omega_{1,2},t)=\int_{t-\tau}^{t+\tau}\mathbf{E}(\mathbf{r},t')\mathrm{e}^{i\omega_{1,2}t'}\mathrm{d}t'$
where $\tau=0.5$~ps. This mimics the response of a spectrally selective
time-resolving detector. The laser was pumped at 300 times threshold
in order to assure bistable regime despite a considerable mismatch
of the mode decay rates ($\kappa_{1}\neq\kappa_{2}$), which is also
the reason why $\Phi_{1}$~and~$\Phi_{2}$ show different on-off
contrast \cite{ourPRA09}.

Numerical simulations demonstrate switching between the two modes
on the scale of several (less than 10) picoseconds (Fig.~\ref{fig:numerical}),
with a decent qualitative agreement between FDTD and coupled-mode
results. Analytical calculations tend to underestimate the switching
times compared to numerics, both in this work and in earlier ones
\cite{ourPRA09,ourPSSMethods07}. One possible reason may be due to
the difference in laser system (4-level in FDTD vs. 2-level in analytics)
but this remains a topic of further studies. The on-off contrast is
at least 40~dB. To estimate the switching energy one should take
into account that the energy transferred to a microstructure is a
fraction~$\rho$ of the externally delivered power. In such lasers,
%
\begin{comment}
the absorbed threshold power was experimentally estimated to be at
200~nW , making
\end{comment}
{} the operating absorbed power $P=60\,\mu\textrm{W}$ and $\rho$~is
around 0.1--0.2. From the simulations, the cavity field enhancement
factor was $f=10$. Thus, for $E_{\text{sw}}=E_{\text{mode}}$ the
estimation for the switching pulse energy is $\rho fP\sigma\simeq$15--30~fJ,
comparable to the earlier accounts \cite{HillNature}. It can be further
reduced, e.g, using coupling optimization \cite{fanAPL09}. The storage
time of the proposed device is theoretically infinite as long as the
laser stays pumped and the in-cavity energy fluctuations remain smaller
than $W_{c}/\sqrt{2}$.

%
\begin{comment}
In summary, we have shown analytically and numerically that bistable
microlasers \cite{ourPRL07,ourPRA09} based on structures within state-of-the-art
fabrication capabilities \cite{AtlasovOE} can be used to design an
optical memory cell capable of low-energy, ultrafast all-optical switching.
The simulated model device shows switching between modes 3 nm apart
from each other by pulses with energy estimated at 15--30~fJ in less
than 10~ps.
\end{comment}
{}

The authors acknowledge helpful assistance of C.~Kremers on numerical
simulations. This work was supported in part by the Deutsche Fouschungsgemeinschaft
(FOR 557).

\end{document}